# A Review of Machine Learning and Computational Methods for Protein Phosphorylation Sites Prediction

Farzaneh Esmaili[1], Mahdi Pourmirzaei[1], Shahin Ramazi[2], Elham Yavari[1]
{f.esmaili, m.poormirzaie, s.ramazi, e.yavari}@modares.ac.ir
[1] Department of Information Technology, Tarbiat Modares University
2 Department of Biophysics, Faculty of Biological Sciences, Tarbiat Modares University

## Abstract

Post-translational modifications (PTMs) have vital roles in extending the functional diversity of proteins and as a result, regulating diverse cellular processes in prokaryotic and eukaryotic organisms. Phosphorylation modification is a vital PTM that occurs in most proteins and plays significant roles in many biological processes. Disorders in the phosphorylation process lead to multiple diseases including neurological disorders and cancers.

At first, this study comprehensively reviewed all databases related to phosphorylation sites (p-sites). Secondly, we introduced all steps regarding dataset creation, data preprocessing and method evaluation in p-sites prediction. Next, we investigated p-sites prediction methods which fall into two computational and Machine Learning (ML) groups. Additionally, it was shown that there are basically two main approaches for p-sites prediction by ML: conventional and End-to-End learning, which were given an overview for both of them. Moreover, this study introduced the most important feature extraction techniques which have mostly been used in ML approaches. Finally, we created three test sets from new proteins related to the 2022th released version of the dbPTM database based on general and human species. After evaluating available online tools on the test sets, results showed that the performance of online tools for p-sites prediction are quite weak on new reported phospho-proteins.

## 1 Introduction

Post-translational modifications (PTMs) are biochemical reactions occurring on a protein after its translation [1,2] which change the regulate physicochemical properties, maturity, and activity of most proteins [3,4]. PTMs include cutting, folding, and ligand-binding and adding a modifying group to one or more amino acids, changing the chemical nature of amino acids [5,6]. In recent years, an increasing volume of PTMs data is available because of improvements in mass spectrometry (MS) based on high-throughput proteomics. Such advancements have revolutionized the study of PTMs [7].

There are more than 600 types of PTMs [8] that affect many aspects of cellular functionalities, such as metabolism, signal transduction, activity, stability, and localization of various proteins [9,10]. Recent studies have shown that each modification leads to a multitude of effects on the structure and therefore, the function of the proteins [11]. PTMs include phosphorylation, glycosylation, ubiquitination, sumoylation, acetylation, succinylation, and nitrosylation as well as numerous others in most cellular activities [9,12–16]. PTMs also have key roles in different biological regulatory mechanisms like metabolic pathways, DNA damage response, transcriptional regulation, signaling pathways, protein–protein interactions, apoptosis, cell death, insulin signaling, immune response, and aging [17,18]. Dysregulation in PTMs is contributed to cancer, diabetes, cardiovascular disease, and neurological disorder [19–24].

Phosphorylation is one of the most important reversible PTMs. This modification was firstly discovered in 1906 by Phoebus Levene in the protein vitellin (Phosvitin) [25]. Phosphorylation occurs covalently by adding a phosphategroup with a −2 charge at physiological PH in Serine, Threonine, Tyrosine, and Histidine residues and removal of the phosphate group via protein phosphatases from the modified site. It is known that protein phosphorylation regulates the activity of various enzymes and receptors including signal pathways [26] and can greatly impact folding, function, stability, and subcellular localization of the protein [25,27,28]. This modification plays key roles in eukaryotic signaling and biological processes including protein synthesis, cell division, signal transduction, DNA repair, environmental stress response, regulation of transcription, apoptosis, cellular motility, immune response, metabolism, cell growth, development, cellular differentiation, and aging [29,30]. In eukaryotes, the phosphorylation process is catalyzed via Protein kinases (PKs) differentially and specifically which each PK only modified a subset of substrates to ensure signaling fidelity [31]. More than one-third of human proteins contain phosphorylation and this modification is regulated by approximately 568 human protein kinases and 156 protein phosphatases [29]. In this sense, phosphorylation is one of the widest spread and most extensively studied protein PTMs and it has a significant role in the control of biological processes like proliferation, differentiation, and apoptosis [29,32]. Site mutations or dysregulation of kinase activity, their hyperactivity, malfunction or overexpression and also, hyper phosphorylation of human proteins is associated with certain disease states such as cancers, Alzheimer’s disease (AD), Parkinson's disease (PD), frontotemporal dementia (FTD), and various pathways involving the immune system [27–29,33]. Thus, identifying kinase-specific p-sites is essential for understanding the regulatory mechanisms of phosphorylation.

Multiple experimental methods are used for the detection assays of protein phosphorylation like: liquid chromatography-tandem mass spectrometry (LC–MS/MS), radioactive chemical labeling, and immunological detection, such as proximity ligation assay (PLA), chromatin immunoprecipitation, and western blotting. Although, the combination of LC–MS/MS method with IP strategy is suitable for detection of p-sites in proteins [34,35]. However, through the use of experimental methods, it is very expensive and challenging to detect p-sites on a large scale. Recently, computational methods for distinguishing PTMs have been extremely appealing to scientists. Researchers use experimental data analysis to train machine learning (ML) methods to predict. Currently, most of the existing computational methods are used to predict the phosphorylation and glycosylation sites [36]. Phosphorylation is highly conserved and is central to the regulation of various cellular processes and has significant effects on the stability of the modified proteins. Figure 1 shows protein phosphorylation scheme.

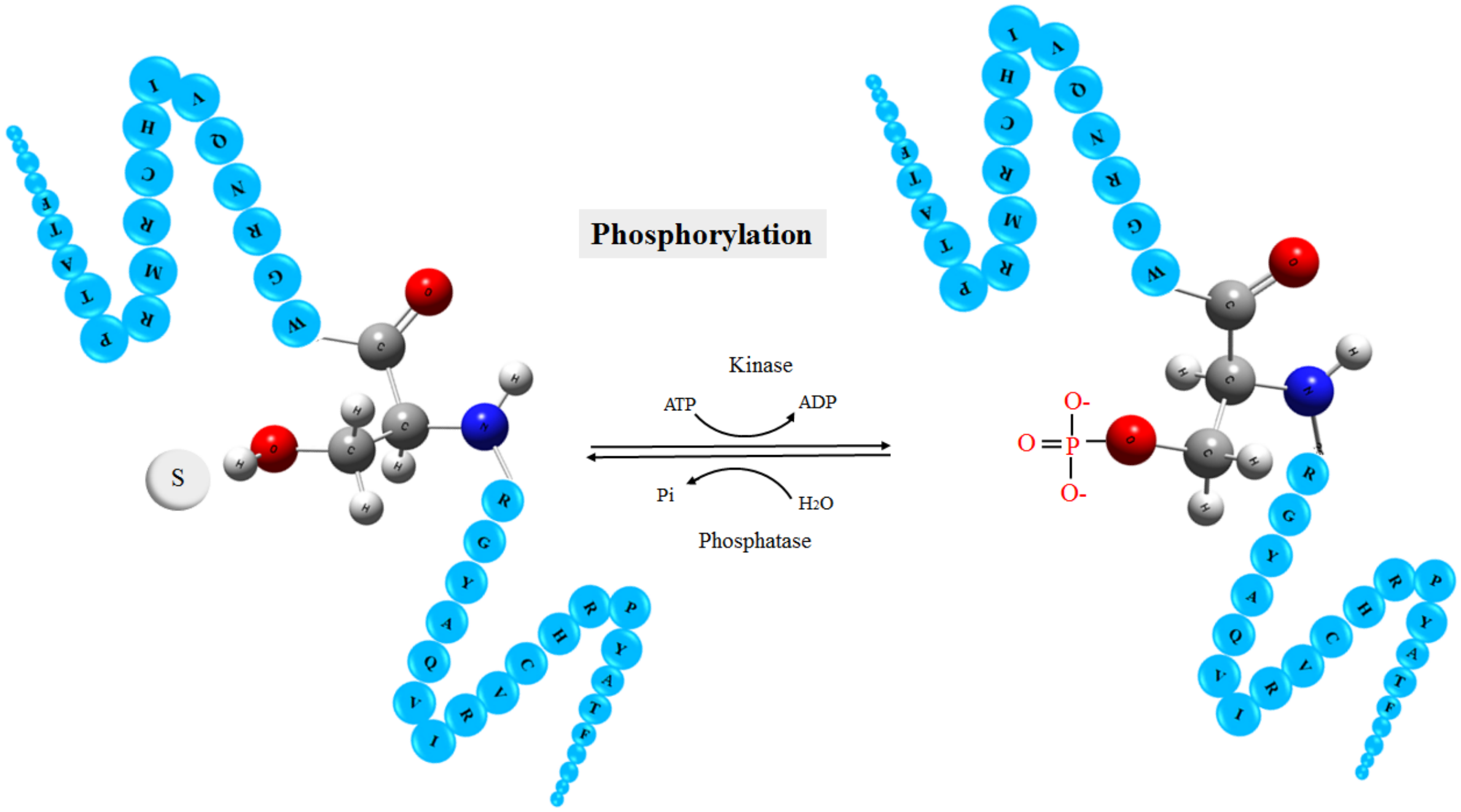

Fig 1: Schema of protein phosphorylation [34].

The number of known phosphorylated sites has grown since 2003, it rose from 2,000 to more than 500,000 known sites in valid databases with using experimental techniques such as tandem mass spectrometry (MS/MS). Furthermore, at least 30% of human Proteomes are regulated by p-sites [37,38]. For instance, in experimental techniques MS/MS was used to recognize 36000 separate p-sites and map the phosphoproteome of nine various mice. It means that MS/MS methods have conceded a lot of information about p-sites. However, there are many technical challenges for using MS/MS approaches and this makes p-sites identification difficult [39]. For example, some proteins like low abundance and proteins which are temporary phosphorylated are missed by using MS/MS approaches [38].

These methods are generally difficult, slow, and costly and needs specialized equipment and knowledge. With the advancement of technology and the emergence of new methods over the last two decades, computational methods and ML have helped the previous methods to identify and predict p-sites.

A Study [34] reviewed PTM tools, resources and related databases and also they investigated the challenges of computational methods. Ramazi and et al. [34], divided 10 types of PTMs into small chemical groups, lipids, and small proteins peptides. They investigated databases and computational approaches for different PTM sites. Shi and et al [40], reviewed 19 available tools for phosphorylation networks. They reported different analyses for their functionality, data sources, performance, network visualization and implementation. Rashid MM and et al. [41], reviewed specified ML methods, main feature selection methods, databases and current online tools for microbial p-sites. They only investigated microbial p-sites and did not mention other p-sites in organisms nonetheless. Also, their work was limited to classical ML methods.

In this study, unlike other previous studies, we investigate all features, databases and methods concerning p-sites prediction. The contribution of this work is summarized as follows:

- Valid PTMs databases contain phosphorylation experimental data were introduced. Then, two most important phosphorylation databases were reviewed in which the number of organisms, p-sites were covered in detail.
- Two main preparing p-sites datasets steps were reviewed which include data collection and data preprocessing. In other words, this study investigated methods for data collection and also introduced the most important and functional approaches for data preprocessing. Additionally, all evaluation metrics which have been used for p-sites prediction were introduced.
- Most common and important feature extraction methods in four types of structural level, sequential, evolutionary and physicochemical property-based were described.
- It was found that there are two machine-learning based approaches exist for p-sites prediction which we divided into conventional and End-to-End learning methods. In the present study, methods of both approaches were reviewed and available online tools of p-site prediction were briefly mentioned.
- In the end, we created three test sets from new proteins related to the 2022nd released version of the dbPTM database. Then, we evaluated and compared available online tools together in different metrics on the three specific test sets.

It should be mentioned that, this article investigated ML, Deep Learning (DL) and computational methods for all p-sites and organisms. Therefore, this paper presents a comprehensive review of p-sites prediction based on ML methods.

## 2 Databases

Databases are constantly evolving due to the advent of technology. Sources for predicting p-sites are expanded by providing accurate information in databases. These databases contain different organisms such as viruses, animals, sapiens, etc. that have been collected manually and experimentally. For instance, the information of HPRD is collected manually and it contains more than 95,000 phosphorylation sites in ~13,000 proteins [42].

Considering different types of PTMs, databases are arranged into specific and general databases. General PTM databases investigate a wide domain of data for different types of PTM modifications. On the other hand, specific databases are constructed based on special types of PTMs like phosphorylation.

Databases such as dbPTM [7], SysPTM [3], SwissProt [43] and HPRD [42] are general databases which cover different types of PTMs and p-site is one of them. Also, EPSD [44], Lymphos2 [45], phospho3d [46], phosphoELM [47] and Regphos [48] are specifically gathered for p-sites.

In the following, two important databases for p-sites are going to introduce. Furthermore, Table 1 investigates both general and specific databases according to their general statistics for p-sites.

## 2.1 EPSD

Eukaryotic Phosphorylation Site Database (EPSD) is one of the most specific, large and comprehensive databases for p-sites which has been updated in 2020. EPSD has updated from two databases of dbPPT [49] and dbPAF [50], which includes roughly ~82,000 p-sites for 20 plants and more than 483,000 p-sites from seven different types of animals and fungi, respectively. Moreover, EPSD collected p-sites in 13 additional databases including PhosphoSitePlus [51], Phospho.ELM [52], UniProt [53], PhosphoPep [54], BioGRID [55], dbPTM, FPD [56], HPRD, MPPD [57], P3DB [58], PHOSIDA [59], PhosPhAt [60], and SysPTM [37]. At all, this database contains ~1,616,800 experimentally known p-sites in approximately 209,300 phosphoproteins of 68 eukaryotes (18 animals, 24 plants, 19 fungi, and 7 protists).

## 2.2 dbPTM

Database post-translation modification (dbPTM) is a general database that integrates PTM's data from 30 databases and ~92,600 research articles. The dbPTM covers 130 types of PTMs in more than 1,000 organisms [34]. The 2022th version of dbPTM [61] has curated more than 2,777,000 PTM sites from 41 published databases and ~82,000 research articles. Figure 2, 3, 4 demonstrates the EPSD database p-sites details.

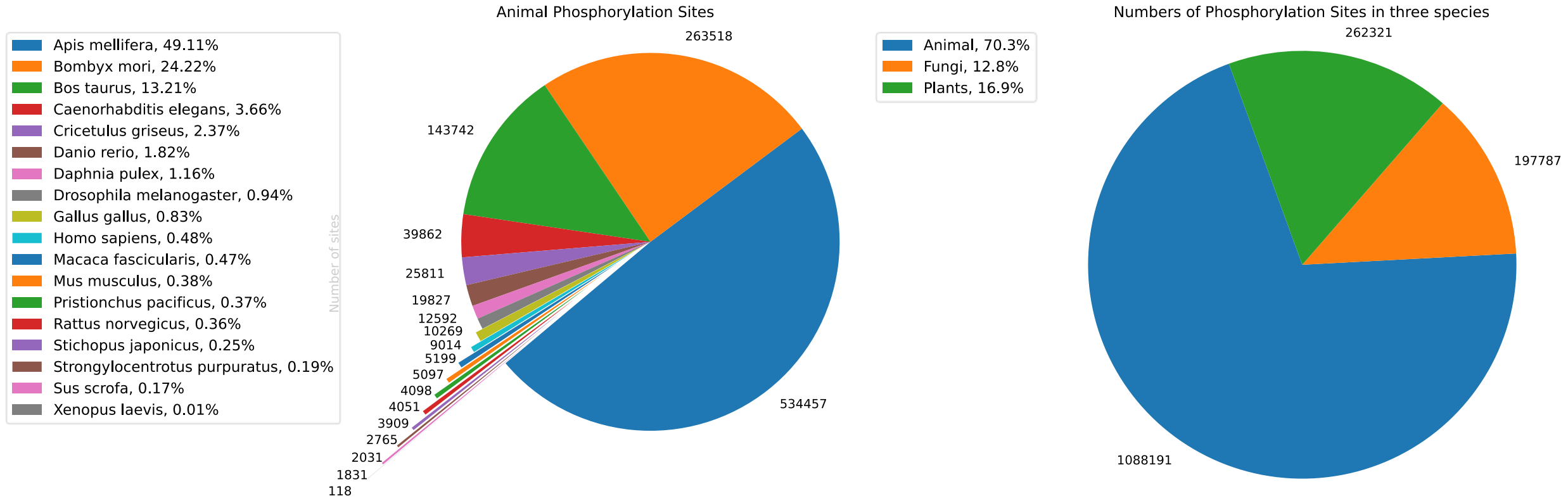

Fig 2: (**Left**) shows the number of p-sites in the animal proteins distributed by different types of animals and (**right**) shows the number of p-sites on proteins related to three species: animal, fungi and plants. All figures are based on EPSD dataset.

Table 1: Contains two general and specific databases which cover number of p-sites and proteins (P). Moreover, it provides useful information about each one. This table is inspired by [34]. * Type of database can be secondary or primary; secondary databases are the integration of other databases. ** Primary databases are independent.

| Type | Acronym | General statistics | | Type of data and database | URL |
|---|---|---|---|---|---|
| | | Number of covered organisms | Number of phosphorylation sites | | |
| **General database** | dbPTM [7] | More than 1,000 organisms | p-sites: ~1,770,000 P: ~557,700 | Experimental and Predicted Secondary* | https://awi.cuhk.edu.cn/dbPTM/ |
| | Phosphosite Plus [62] | 26 organisms | p-sites: ~240,000 P: ~20,200 | Experimental Primary | https://www.phosphosite.org |
| | PTMCode v2 [63] | 19 organisms | p-sites: ~316,500 P: ~45,300 | Experimental Secondary | http://ptmcode.embl.de |
| | qPTM [64] | Human | p-sites: ~199,000 P: ~18,402 | Experimental Secondary | http://qptm.omicsbio.info/ |
| | YAAM [65] | Saccharomyces cerevisiae | p-sites: ~3,900 P: ~680 | Experimental Secondary | http://yaam.ifc.unam.mx |
| | HPRD [42] | Human | p-sites: ~1,100 P: ~30,000 | Experimental Primary | http://www.hprd.org |
| | PHOSIDA [59] | 9 organisms | p-sites: ~70,000 P: ~28,700 | Experimental Secondary | http://www.phosida.com |
| | PTM-SD [1] | 7 model organisms | p-sites: ~1,600 P: ~842 | Experimental Secondary | http://www.dsimb.inserm.fr/dsimb_tools/PTM-SD |
| | SysPTM [3] | 6 organisms | p-sites: ~353,000 P: ~53,200 | Experimental Secondary | http://lifecenter.sgst.cn/SysPTM/ |
| **Phosphorylation databases** | EPSD [44] | 68 organisms | p-sites:~1,616,800 P: ~209,300 | Experimental Secondary | http://epsd.biocuckoo.cn |
| | PhosphoNET [66] | Human | p-sites: ~966,000 P: ~20,000 | Experimental and Predicted Secondary | http://www.phosphonet.ca |
| | RegPhos [48] | Human, mouse and rat | p-sites: ~113,000 P: ~18,700 | Experimental and Predicted Secondary | http://140.138.144.141/~RegPhos |
| | Phospho.ELM [47] | Mainly model organims | p-sites: ~42,500 P: ~8,600 | Experimental Secondary | http://phospho.elm.eu.org |
| | Phospho3D [46] | Mainly model organisms | p-sites: ~42,500 P: ~8,700 | Experimental Secondary | http://www.phospho3d.org |
| | dbPSP [67] | 200 prokaryotic organisms | p-sites: ~19,300 P: ~8,600 | Experimental Secondary | http://dbpsp.biocuckoo.cn/indExp.php |
| | pTestis [68] | Mouse | p-sites: ~17,800 P: ~3,900 | Experimental and Predicted Secondary | http://ptestis.biocuckoo.org |
| | LymPHOS [45] | Human Mouse | p-sites: ~18,300 P: ~4,900 | Experimental and Predicted Primary | https://www.lymphos.org |
| | P3DB [58] | 45 plant organisms | p-sites: ~220,000 P: ~57,000 | Experimental and Predicted | http://www.p3db.org |

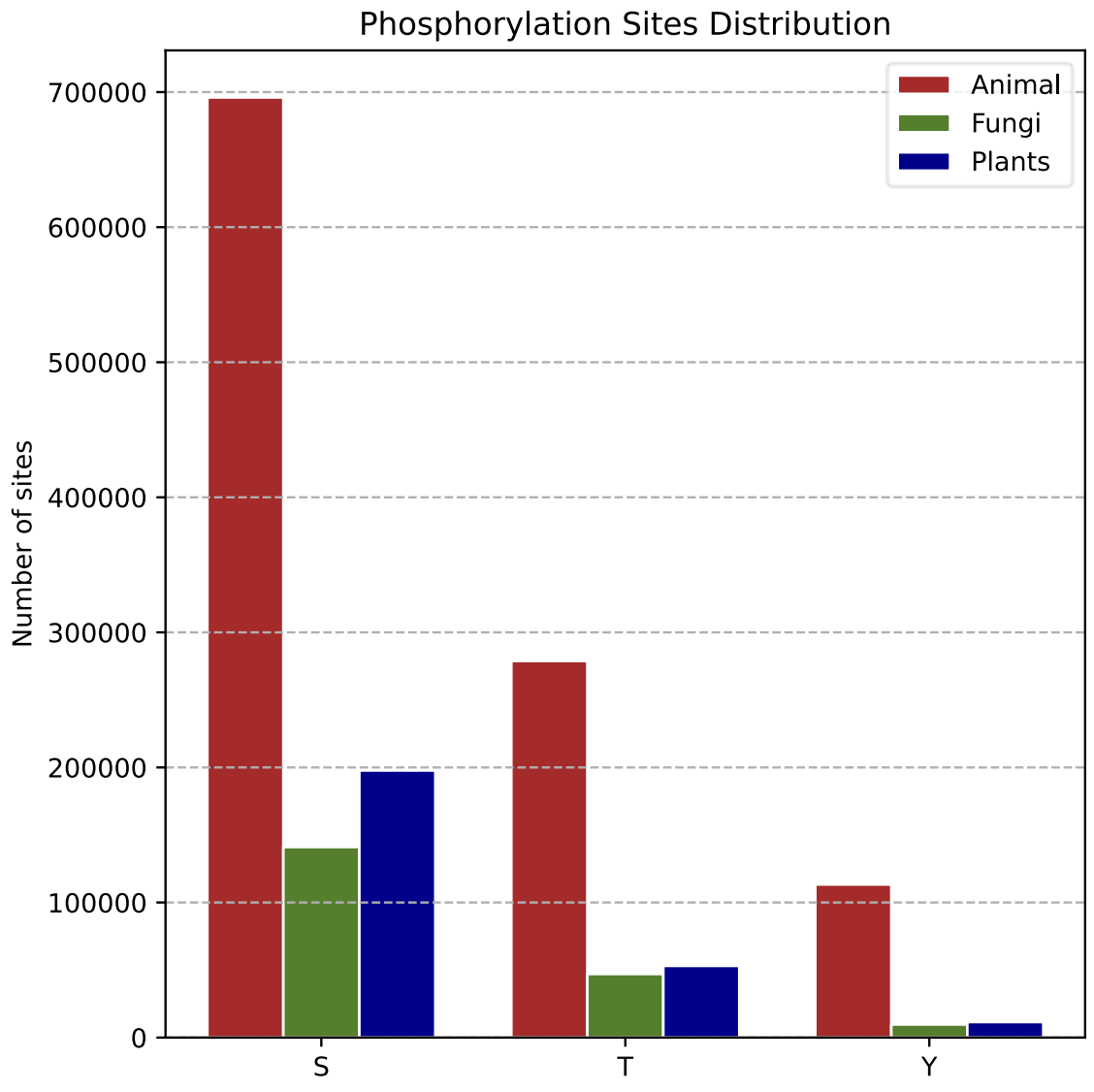

Fig 3: Number of S, T, and Y in proteins related to animal, plants, and fungi organisms in EPSD database.

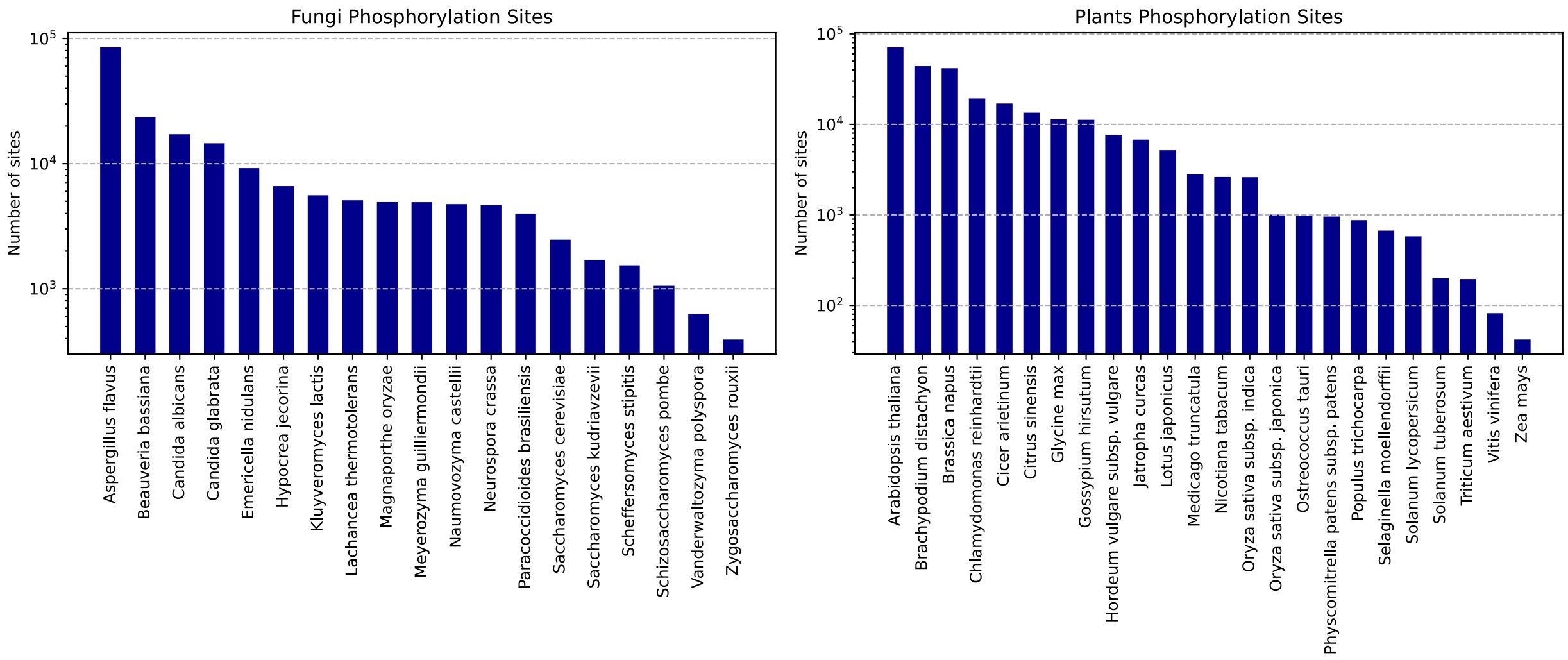

Fig 4: Distribution of p-sites in EPSD database in log scale for (**left**) fungi proteins and (**right**) plants proteins.

## 2.3 Identifying driver mutations and their effects on p-site proteins

Phosphorylation is involved in a wide range of aspects of cellular organization and signaling pathways associated with disease. Various studies have demonstrated that p-sites are evolutionarily constrained in human genomes, as well as prevalent in cancer driver mutations and causal variants of inherited disease. Thus, phosphorylation information and detection of its functional are useful for interpreting genetic variation, genotype-phenotype associations, and molecular disease and their treatment [69].

DNA single nucleotide variants (SNVs) are caused by a single nucleotide change, which is the most common type of sequence changes. Genetic variation of p-sites via SVNs can directly have an effect on modifying target residues or indirectly by modifying the consensus binding sequences (i.e., short linear motifs) located in the flanking sequences of phosphorylated residues. As a result, this can change signaling networks by making, changing, and disrupting the p-sites [70]. There have been reports of phosphorylation-related SNVs that disrupt existing sites and create new sites, disrupting kinase-substrate interactions and causing disease phenotypes. A major challenge facing biomedical research is the identification of genotype phenotype associations, molecular mechanisms, and cancer driver mutations [69].

There are various databases with a useful list of genome variants in p-sites and other PTMs sites. However, they do not provide methods that automatically predict how mutations on p-sites and other protein sites will affect kinase binding. Thus, databases and updated tools are required to interpret rapidly increasing genomic and phosphoproteomic data to interpret signaling networks. We are briefly going to describe ActiveDriverDB database as well as MIMP and PTMsnp tools in this field.

The ActiveDriverDB is a web database which was designed to understand how protein coding varies in the human genomes. The ActiveDriverDB database contains more than 260,000 experimentally identified PTMs sites in the human proteome using public databases like PhosphoSitePlus, UniProt, Phospho.ELM, and HPRD which contains ~149,300 p-sites. As evidenced in the ActiveDriverDB database, changes in target amino acids substitutions in p- sites influence the creation of pathogenic disease mutations, somatic mutations in cancer genomes, and germline variants in humans. Additionally, the ActiveDriverDB database contains phosphoproteomics data reflecting the cellular response to SARSCoV-2 infection, which can be used to predict the impact of human genetic variation on COVID-19 infection and disease course [70].

Mutation impact on phosphorylation (MIMP) (http://mimp.baderlab.org/) is an online tool for predicting the impact of missense SNVs on kinase-substrate interactions. MIMP analyzes kinase sequence specificities and predicts whether SNVs disrupt existing p-sites or create new sites. This helps discover mutations that modify protein function by altering kinase networks and provides insight into disease biology and therapy development [69].

PTMsnp is an online tool concerning identifying driver genetic mutations aiming at PTM sites in proteins across different cohorts from TCGA by using a Bayesian hierarchical model. There are more than 411,500 modification sites in PTMsnp from 33 different types of PTMs and 1,776,800 mutation sites from 33 types of cancer. The web server detects proteins with higher frequency of PTM-specific mutations in the motif region, considered to be the key targets in human disease development [71].

# 3. Data gathering and preprocessing

In this section, we are going to describe steps concerning creating and preprocessing datasets before p-sites prediction. In the last decade, due to the importance of phosphorylation in understanding biological systems of proteins and in guiding basic biomedical drug design, research on phosphorylation has boomed. Several experimental methods are used to identify p-sites in a large number of phosphorylation examples with high accuracy but many of them are labor-intensive and time-consuming. Therefore, low-cost and fast, computational and ML methods have become popular to overcome the problems associated with experimental methods [72]. In order to build the dataset for p-sites prediction, all verified data from multiple databases are considered. Mainly, there are two main steps to create a dataset [72].

1. Data collection
2. Data preprocessing

**Positive data collection**: The S, T, and Y amino acids as p-sites and the positive samples are usually compiled from the aforementioned databases (e.g., EPSD and dbPTM).

**Negative data collection**: S, T, and Y amino acids existing in experimental peptides without any phospho-groups are considered as non-p-sites or negative samples.

Data gathering is the most challenging when selecting the negative dataset and thus, there are two major strategies accessible to choose the negative samples:

- From phosphoproteins, the negative random samples of the target residue that did not undergo the phosphorylation modifications are selected.
- From non-phosphoproteins with none of their target residues (S, T, and Y) that have undergone specific phosphorylation, (based on experimental evidence) are selected as the negative set.

## 3.1 Data preprocessing

After constructing the primary positive and negative datasets, one important task is removing inconsistent/redundant samples to gain a more reliable dataset. This step varies from study to study. One that can distinguish three main policies in the literature for removing inconsistent/redundant proteins is:

1. Removing redundant phosphoproteins.
2. Removing identical subsequences within the positive and negative sets.
3. Removing identical subsequences between the positive and negative datasets.

The Cluster Database at high identity with tolerance (CD-HIT) program is designed to reduce homology and filter out similar sequences. According to different phosphorylation prediction studies [72–75], a threshold of identity is to consider a pair of sequences to be similar/redundant differs, and this threshold is considered to range from 30% to 60% in many phosphorylation prediction studies [76].

## 3.2 Class imbalanced problem

There is a common problem in some ML datasets that happens when the distribution ratios of classes differ. Thus, the dataset is imbalanced and we encounter a class imbalance problem. In other words, a dataset that has unequal samples in classes is imbalanced. This is not a problem when the difference is not that much. Nevertheless, when one or more classes are infrequent, many models do not work too well at identifying the minority classes. For example, in p-site prediction, mostly, preprocessed phosphorylation datasets are imbalanced and the number of the negative samples is much greater than positive samples. Figure 5 shows the preprocessing flow with balancing data.

In the following, three most used approaches to deal with class imbalanced problems are introduced:

**Up sampling**: It generates additional data for minority classes either by making copies of the minimum class or by creating synthetic data which can represent samples of minimum classes**.**

**Down sampling**: It removes data from the majority class either by picking them up randomly or using other approaches of selecting appropriately to handle the issue. Usually, this has been a basic method in p-site prediction, which balanced negative samples by randomly selecting them to become equal to positive samples in numbers.

**Customize loss function**: This is a group/series of methods to deal with imbalance problems in ML. These sorts of methods try to customize the loss function by assigning larger weights to minority classes in order to overcome the issue. However, recently, with the emerging Deep Neural Networks (DNN), a big training set has become crucial and important. Furthermore, customized losses demonstrated better performance and have attracted more attention than up sampling and down sampling approaches.

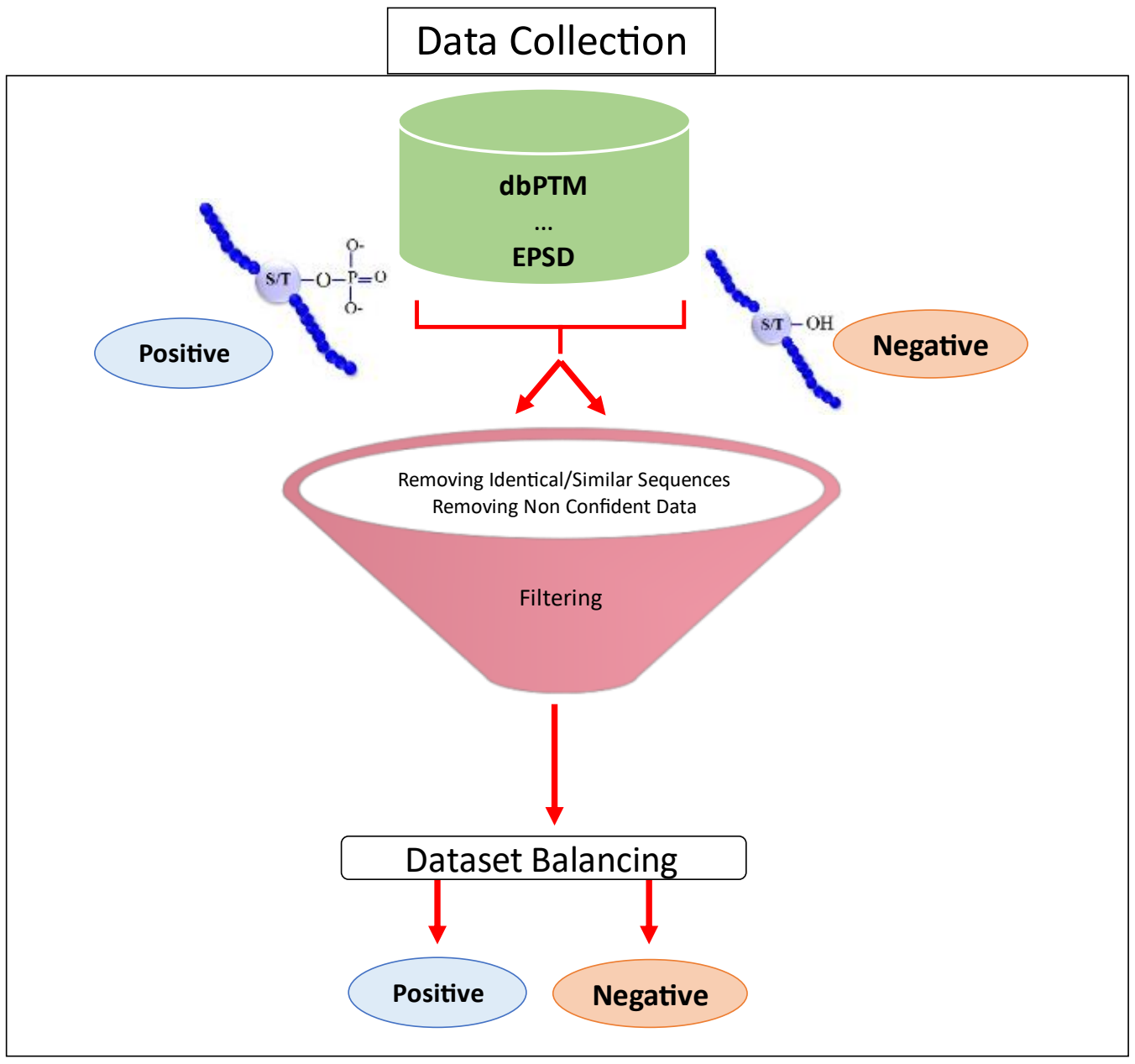

Fig 5: Data preprocessing flow which includes the balancing step.

# 4 Evaluation

The evaluation metrics of protein p-sites are classified into five methods using different attributes: Accuracy (ACC), Sensitivity (SN), Specificity (SP) the Matthews Coefficients of Correlation (MCC), and the area under the ROC curve (AUC). These metrics are evaluated with a confusion matrix that compares the actual target values with those predicted by a model. The number of rows and columns in this matrix depend on the number of classes. From the confusion matrix we end up with four values [34,77]:

**True positive (TP)**: Indicates the number of positive samples that the model classified correctly.

**False Positive (FP):** Indicates the number of negative samples that the model classified incorrectly.

**True Negative (TN):** Indicates the number of negative samples that the model classified correctly.

**False Negative (FN):** Indicates the number of positive samples that the algorithm classified incorrectly.

ACC is the percentage of correct predictions. This metric is defined in Equation 1 as the ratio of both true positive (TP) and true negative (TN) to the total number of cases examined which are false negative (FN) and false positive (FP).

$$Accuracy = \frac{TP + TN}{TP + TN + FP + FN} \tag{1}$$

The SN or Recall is the proportion of true positive prediction to all positive cases (Equation 2).

$$Recall = \frac{TP}{TP + FN} \tag{2}$$

The SP is defined in Equation 3. It calculates the proportion of samples which got predicted truly to all negative samples.

$$Specificity = \frac{TN}{TN + FP} \tag{3}$$

The Precision metric is defined in Equation 4. It calculates the proportion of true positive samples to all cases that were predicted as positive.

$$Precision = \frac{TP}{TP + FP} \tag{4}$$

The F1-score is a combination of Precision and Recall which is defined as in Equation 5. This metric facilitates the process. It can be used to compare the performance of methods with a single number.

$$F1 = \frac{2 \times Precision \times Recall}{Precision + Recall} \tag{5}$$

Two SN and SP measures are used to plot the ROC curve and AUC is used to determine the model performance. Furthermore, for binary classification, there is a more elegant solution: Treat the true class and the predicted class as two (binary) variables, and compute their Correlation Coefficient (Equation 6). The higher the correlation between true and predicted values, the better the prediction. This is the rechristened MCC when applied to classifiers.

$$MCC = \frac{TP \times TN - FP \times FN}{\sqrt{(TP + FN)(TP + FP)(TN + FN)(TN + FP)}} \tag{6}$$

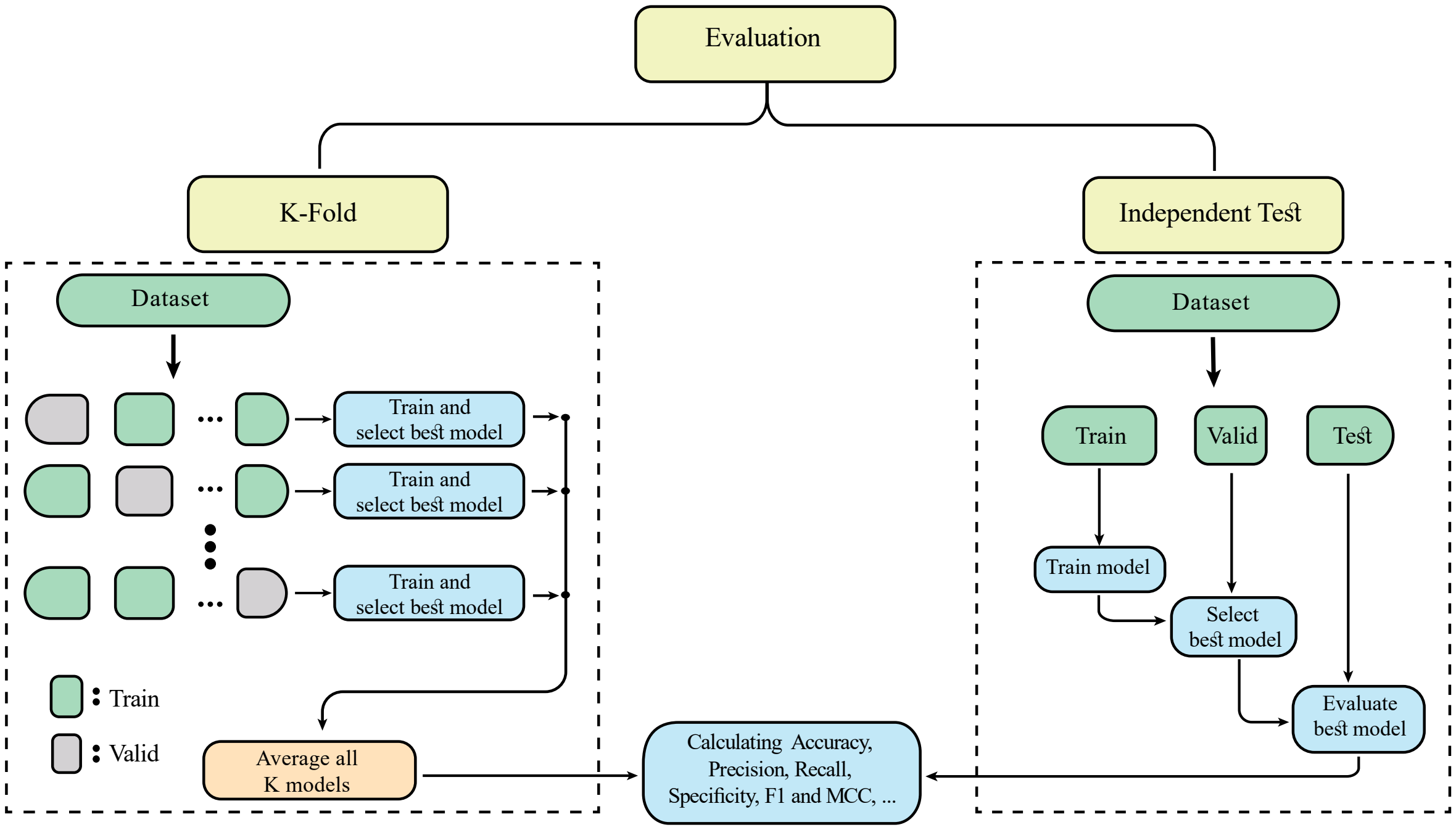

Fig 6: The evaluation step can be done by two methods: K-fold cross validation and independent test. Independent test method sometimes is called "Train-Test" or "Train-Valid-Test" as well.

## 4.1 Model evaluation

Basically, model evaluation is divided into two methods: Independent test (Train-Test) and K-fold cross-validation. In the first one, a dataset splits into two sets: A Train-Valid and a test set. Then, the Train-Valid set splits into two subsets again: a train set and a valid set. The basic procedure is that the train set is used to train models and the valid set is used for the evaluation of the trained models. After selecting the best model with respect to the valid set result, we need to evaluate it on the test set. If the valid set evaluation results are different from the train set, it means that the model is overfitted on the train set. At the end, we should report the test set and there shouldn't be much difference between the valid set and the test set results (Figure 6).

On the other hand, K-fold cross-validation is a resampling method used to evaluate ML models on a limited data sample. The procedure has a single parameter called k that refers to the number of groups that a given data sample is to be split into. As such, the procedure is often called k-fold cross-validation. Specific values for k can be chosen. Considering the scenario of 5-Fold cross-validation (k=5). Here, the data set is split into 5 folds. In the first iteration, the first fold is used to evaluate the model and the rest are used to train the model. In the second iteration, 2-fold is used as the valid set while the rest serve as the training set. This process is repeated until each fold of the 5 folds has been used as the valid set. Each sample is given the opportunity to be used in the hold-out set one time and used to train the model k-1 times. The k-fold Cross-Validation is usually used when the amount of Train-Valid data is limited. On the contrary, when dealing with huge amounts of data, we do not need to have a big valid set. In other words, the proportion of Train-Valid split sometimes can go below 1% for the valid set. This approach is mostly used when massive amounts of data are accessible. But in low data regimes, they usually split with proportions of 70%-30%.

Need to mention that there is another evaluation method exists named Jack-knife validation test [78]. Jack-knife validation test (Sometimes called leave-one-out cross-validation test) is the most objective validation method and provides unique results for a dataset in which one sample is selected to be the test data and the rest are the training data. This procedure will be repeated N times in a data set with N samples which could be expensive for big datasets [79]. This evaluation technique has been used rarely for p-site prediction.

In summary, K-fold should be used in low data regimes and an independent method with a small percentage of valid set should be used when we have access to lots of data.

# 5 Methods for predicting phosphorylation sites

In the following sections, we are going to review methods of p-sites classification by dividing them into two main categories: computational and ML. Likewise, ML methods are also divided into two approaches: conventional and End-to-End learning methods. These two are going to be described below in the following paragraphs.

## 5.1 Computational methods

Innovative methods based on statistical approaches have been used in many studies. In Schwartz and Gygi [80], a statistically repetitive method, a set of phosphorylated peptide sequences to extract the patterns and a set of peptide sequences to evaluate the predictions were used. They mapped two sets of sequences to the position-weight matrix so that in the matrices, the number of repetitions of each residue was determined from 6 positions higher to 6 positions lower than each p-site (it means their window size for each peptide is 13 amino acids long). Then they formed a binary matrix based on these two matrices. This final matrix indicates the probability of observing a specific residue around a p-sites by examining this matrix and comparing it with other p-sites. They extracted some of the residues found around most p-sites and used them to predict new p-sites.

Chen et al. [81] presented a new method for predicting p-sites by collecting four background datasets including phosphorylated and non-phosphorylated sequences. They chose a given length of 13 for windows around p-sites. Initially, they formed weight-position matrices; then, they extracted patterns. By scoring those patterns and deleting some of them, they finally reported a series of patterns as the output during an iterative cycle.

He et al. [82], showed that the number of patterns to be examined around each positions are growing exponentially based on the length of the window. They refer to two developed algorithms to find phosphorylation patterns, named the Motif-X and the MoDL algorithms. They supposed that these algorithms do not detect all patterns and some patterns remain hidden from biologists. Thus, they introduced a new algorithm called Motif-ALL to discover and report all possible patterns based on previous algorithms.

There have been a family of algorithms called Group-based Prediction System (GPS) for many years in computational methods [83–89]. In 2004, an algorithm was developed, group-based p-site predicting and scoring 1.0, based on the hypothesis that similar short peptides exhibit similar biological functions. Likewise, the algorithm was refined and constructed an online service of GPS 1.1, which could predict p-sites for 71 PK clusters. Then, GPS 2.0 and 2.1 were presented in which two methods matrix mutation (MaM) and motif length selection (MLS) were designed to improve the prediction accuracy, whereas the scoring strategy was not changed. Consequently, GPS 2.2, 3.0, 4.0 and 5.0 algorithms were developed which are used for the prediction of PTM sites other than p-site [31].

## 5.2 Machine learning methods

Most of algorithms used for phosphorylation prediction are based on ML algorithms. Moreover, by explosions of DL method in the early 2010s, ML gets popular even more than before. ML is generally the ability of machines to do actions based on prior knowledge and experience [90]. There are more than 40 different methods for predicting p-sites and many of which use various ML methods including Logistic Regression (LR), Support Machine Vector (SVM), Random Forest (RF), and K-Nearest Neighbor (KNN) [72].

In general, there are two main directions to predict Phosphorylation in ML: conventional and End-to-End learning methods. The conventional approach stands for using ML algorithms as a part of solving a solution besides other steps in pipeline designs such as feature extraction and hand-feature engineering. On the other hand, the End-to-End learning approach stands for the new wave of DL algorithms by which hand-craft features removed from the solution. In other words, it refers to training a possibly complex learning system represented by a single model (specifically a DNN) that represents the complete target system, bypassing the intermediate layers usually present in traditional pipeline designs.

### 5.2.1 Feature extraction

In protein phosphorylation prediction, various types of conventional approaches have been studied. The most common of these approaches use different methods as feature extraction [91]. In this paper, we reviewed 20 feature extraction techniques which are extracted according to physicochemical, sequences, evolutionary and structural features and convert each sequence into numerical vectors for feeding them as an input to be classified by an algorithm. We have tried to introduce important and practical methods of feature extraction in this paper, but it is clear that there may be several techniques for information extraction. In the following, the most important ones are going to explain.

#### 5.2.1.1 Physicochemical property-based features

**Encoding based on grouped weight ( EBGW):** EBGW divides 20 amino acid into 7 categories based on their hydrophobicity and charge characteristics [92,93]. For each group $H_i$ (i =1, 2, 3) a 25-dimensional array Si (i = 1, 2, 3) of the same element in the segment should be generated. If the amino acid at that position is belonged to the $H_i$ group, the element in the array will set to 1, otherwise, it will set to 0. Each array will be divided into sub-arrays (*J*-ones), which represent as *D(j)*. This value can be taken from cutting the main $S_i$ from the first window of *len(D(j))* which defines as Equation 7:

$$len(D(j)) = int\left(\frac{j * L}{J}\right) j = 1, 2, \ldots, J \quad , \qquad L = length\ of\ segments \tag{7}$$

For each group of $H_i$, a vector with length of J based on its sub arrays should be defined in which the j-th element of $X_i^{(j)}$, is calculated based on Equation 8:

$$X_i^{(j)} = \frac{Sum\ (D(j))}{len\ (D(j))} \tag{8}$$

**Amino Acid Index (AAINDEX):** It extracts features based on amino acid indices from AAINDEX database. This database was used for prediction different types of PTM and also p-sites [94]. According to physicochemical and biological properties, hydrophobicity, polarity, polarizability, solvent /hydration potential, accessibility reduction ratio, net charge index of side chains, molecular weight, PK-N, PK-C, melting point, optical rotation, entropy of formation, heat capacity and absolute entropy, each amino acid in each position is represented by the mentioned 14 values [92,95].

**Average Accumulated Hydrophobicity (ACH)**: ACH quantifies the tendency of amino acids surrounding S, T, or Y residues to be exposed to solvent [96]. ACH is computed by averaging the cumulative hydrophobicity indices around the p-site for different sliding windows. It should be mentioned that every site is located in the center of the sliding windows [97,98].

**Encoding scheme Based on Attribute Grouping (EBAG)**: Encoding scheme of protein sequences was considered hydrophobicity attribute and divided amino acid residues into 4 classes based on physicochemical property. The hydrophobic class c1={A, F, G, I, L, M, P, V, W} polar class c2={C, N, Q, S, T, Y} acidic class c3={D, E} and basic class c4={H, K, R} were discussed [99,100].

**Overlapping Properties (OP)**: OP clusters each protein based on their chemical attributes. Each amino acid is classified into 10 physicochemical properties: polar, positive, negative, charged, hydrophobic, aliphatic, aromatic, small, tiny and proline [98].

**Pseudo amino acid compositions (PseAAC):** This feature is firstly defined by Chou and et.al [101] for coding proteins. They proposed sequence order and physicochemical information in protein sequences. For more details refer to [102–105]

### 5.2.1.2 Sequence-based features

**Quasi-sequence order (QSO):** It reflects the occurrence of amino acids based on two distance matrices: Physicochemical and chemical distance matrices [92]. Most physicochemical properties are hydrophobicity, hydrophilicity, polarity and side-chain volume. Features are derived from a distance matrix created by computing the distance between each pair of the 20 amino acids [98,106,107]. This feature was originally proposed by Chou and et.al [101]. For more detail refer to [101,108].

**Numerical representation for amino acids:** It converts each character of amino acids into numerical numbers as mapping them in alphabetic order from 1 to 20 and dummy amino acid X represents 21 [92].

**Binary encoding of amino acids (BINA):** BINA represents each amino acid as 21-dimensional binary vectors, which encode 1 for the target amino acid and 0 for the residues (other 20 amino acids). For example, alanine (‘A’) is shown as 100000000000000000000 [92].

**LOGO**: This feature is defined with calculating the occurrence of amino acid frequencies and encoding them in a sequence with Two Sample Logo program [92].

**Position Weight Amino Acid composition (PWAA)**: Position information of each amino acid is another key point that shall be considered in feature extraction. PWAA can reveal sequence order information around P, S, and Y residues [109]. PWAA can be declared from Equation 9.

$$C_i = \frac{1}{L(L+1)} \sum_{j=-L}^{L} x_{i,j} \left( j + \frac{|j|}{L} \right), \ j = -L, \dots, L$$

(9)

**Composition of K-Spaced Amino Acid Pairs (CKAAP)**: The encoding of CKAAP is pretty easy, which can directly be calculated from the sequence pieces of p-sites and non-p-sites. CKSAAP is a critical encoding scheme feature selection in lots of prediction tasks, especially in representing short sequence residues in protein sequence or subsequence. A subsequence may contain 400 types (AxA, AxC, AxD, …, OxO) of K-spaced amino acid pairs (i.e. the pairs separated by K other amino acids) [110]. CKAAP equation was proposed as it follows in Equation 10 [74].

$$f_{i,j} = \frac{Num\left(A_i A_j\right)}{L - K - 1} \ i, j = 1,2, \dots 21$$

(10)

**Amino acid compositions (AAC)** : This method is the most common one in feature selection methods, which calculates each amino acid's frequency in protein sequences [111]. AAC is a simple way to extract information from a sequence of amino acids according by encoding them into 20 bits. Also, it depends on the window size of each protein which they encode with a unique 20-bit of one amino acids [98].

Lin and et al. [111] proposed AAC equation as Equation 11.

$$v_i = \frac{c_i}{len(seq)} \ i = 1, \dots., 20$$

(11)

### 5.2.1.3 Evolutionary-based features

**K-Nearest Neighbor (KNN):** The most popular feature selection method which is used in various ML problems especially in PTM and phosphorylation classification is KNN. It classifies sequences based on their distance. The algorithm classifies sequences by looking at k of nearest neighbor sequences by finding out majority votes from nearest neighbors that have similar attributes and the shortest distance as those used to map the items [112].

**Position-Specific Scoring Matrix based transformation (PSSM):** Position-Specific Scoring Matrix (PSSM) encoded the evolutionary data of a protein and PSSM profile is informative and useful for a number of biological classification problems. According to Equation 12, the PSSM matrix in a protein with a sequence of length L is a matrix with L * 20 dimensions. In the matrix, each row represents an amino acid in the protein sequence, and the columns represent the 20 amino acids in proteins.

### 5.2.1.4 Structural-based features

**Protein Disorder Features** (**DF**): All PTM modifications include p-sites located within disorder positions [113]. Protein disorders were used as features in many works such as Iakoucheva and et al. [114], which created the phosphorylation predictor-DISPHO and used protein disorders as features.

In another work [97], disorder information was extracted using VSL2B [115] and the disorder scores for both positive and negative datasets were calculated and the average scores were used for different window sizes as final features.

**Shannon Entropy (H)**: Entropy in information theory is a numerical measure of the amount of information or randomness of a random variable. To be more precise, the entropy of a random variable is the average value (Expected value) of the amount of information obtained from observing it. It means, when the entropy of a random variable is high, we have more ambiguity about that random variable. Therefore, by observing the definite result of that random variable, more information is obtained, so when the entropy of a random variable is high, more information will be obtained from its definite observation [116]. In science and engineering in general, entropy is a measure of the degree of ambiguity or disorders [117]. Claude Shannon, in his revolutionary paper A Mathematical Theory of Communication in 1948, introduced Shannon entropy and became the founder of information theory.

**Relative Entropy (RE)**: It is known as Kullback Leibler which is aggregated entropies for more than 20 sites in proteins[118].

**Information gain (IG):** IG can be computed by subtracting RE from entropy. It can measure the transformation of information from the background or random state to the state influenced by the class whether the sequence is positive or negative [98]. IG is calculated by Equation 12:

$$IG = H - RE \tag{12}$$

**Accessible Surface Area (ASA)**: Accessible surface area or solvent-accessible surface area is a biomolecule surface which can access the solvent. Amino acids can be both exposed and hidden based on 3-dimensional structure. Hidden amino acids are not hidden amino acids do not undergo PTM due to they do not interact with enzymes [98]. On the other hand, a study proposed a Rvp-net algorithm which is used to extract ASA features from protein sequences before converting them into sliding windows [119].

## 5.2.2 Conventional machine learning approach

Once the features have been extracted, models that need to predict p-sites should be developed. Thus, the ML-based methods were reviewed in this section. One of the most popular methods which is currently used for predicting sites is SVM [97,111,120]. SVMs are a set of points in the n-dimensional space of data that define the boundaries of categories. The data is bounded and categorized based on them, and by moving one unit of the data, category output may change. The SVM is a kind of maximum margin classifier. The maximum margin classifier has a simple function, in which data is separated by a hyperplane; provided they have the highest margin over the data. But the maximum margin classifier cannot be used on all datasets, because the data must be linearly separable, which is not the case. If the data cannot be separated linearly, it can be transferred to a higher dimension. By converting and mapping data to a higher dimension, they are transformed from nonlinear separators to linear separators. This is because as the dimension increases, the data becomes more fragmented and open, and this dimension increase can be continued until they become linearly separated [121,122]. The SVM is widely used in bioinformatics, especially in PTM issues. By this method, the protein sequences are filtered and converted to feature vectors of constant length and subjected to be classified into correct classes in which one is considered as phosphorylated and zero is non-phosphorylated [97,109,123].

RF is one of the well-known and important ML algorithms that are used in many issues in the field of bioinformatics and life sciences. RF is a supervised learning algorithm. As the name implies, this algorithm builds a forest randomly. Forests are actually a group of decision trees. RF makes several decision trees and merges them together to make more accurate and stable predictions [124]. Figure 7 demonstrates the procedure of feature extraction methods and also, Figure 8 shows the process of ML conventional methods.

As we mentioned, in the last few years, kinases specific methods were considered because in general, some protein prediction sites remain unexplored and kinases give us a lightening way to find them. NetPhos [125] and NetPhosK [126] both used DNN based on consensus sequences and combined it with mass spectrometry experimental methods. These algorithms are specific to the kinases family. In the Quokka framework [30] they used LR to classify 43 S/T and 22 Y kinases family sites. Another method [120] proposed a classification method based on SVM and used consensus sequence structure as features for four kinases groups and families. The best accuracies which the model could predict were from 83 to 95% at the kinase family level, and 76–91% at the kinase groups. Liu W and et.al [127] proposed a method for four kinases families based on RF which extracts features with Auto Covariance (AC) transform and seven physicochemical properties performed over 90% accuracy.

To recognize protein p-sites in universal proteins, research by Huang S-Y and .et al [109] proposed a method based on SVM in viruses. They used EBAG and PWAA features for extracting physicochemical and sequence information of viral proteins around p-sites. They used a 10-fold cross-validation for different window sizes from 15-27 lengths. They got the best results for window size 23 with an accuracy of 88.8%, 95.2%, and 97.1% for Phosphoserine, Phosphothreonine and Phosphotyrosine respectively. They also showed the influence of using different features. Their model improved almost 15% when they used the combination of two EBAG and PWAA features.

Furthermore, Lin et al. [111] used KNN, AF and CKAAP as features and combined different features together to feed it to their model to investigate best features. The combination of AF and CKAAP provided best accuracy for their SVM model. They believed SVM could classify rice protein universal p-sites. Their work was named Rice Phospho 1.0 which achieved 82% accuracy.

A paper proposed Granularity Support Vector Machine (GSVM) for predicting universal p-sites [97]. They used KNN, AF, DF, and ACH of every phosphorylation position for building training samples. To partition data into high-dimensional feature spaces, they used kernel fuzzy c-means clustering and by applying this method, they tried to extract features for dataset representation in a sample space. The method was applied on plants and animal dataset types and could achieve 80% and 85% accuracy, respectively.

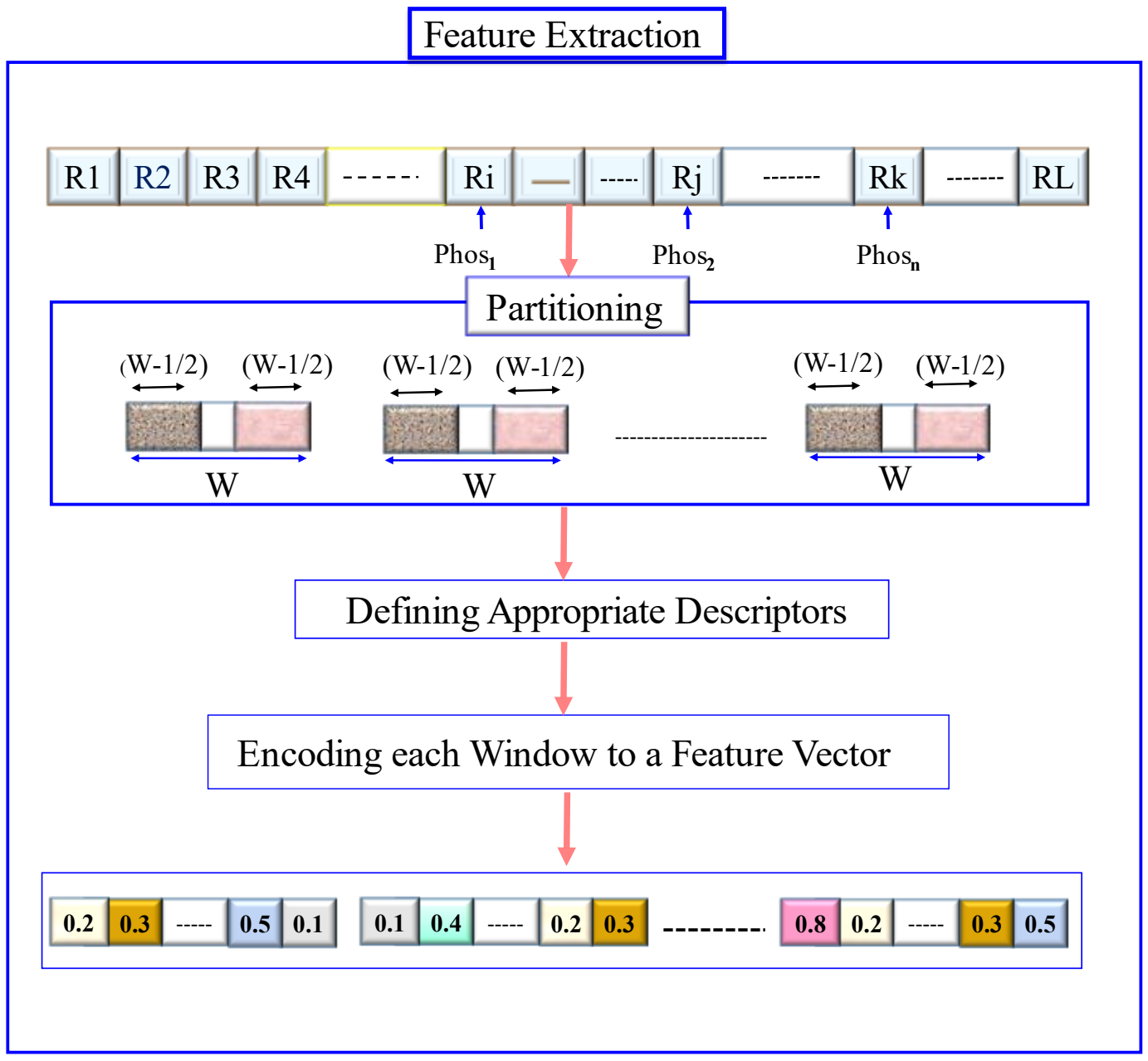

Fig 7: A common procedure in feature extraction stage.

By Phospred-RF method, Banerjee and et al. [128] used information extracted from PSSW and trained individuals RF with odd window sizes from 9-25 amino acids. They got approximately 70% accuracy for 26 protein sequences. RF-phos-1.0 transformed each amino acid to vectors by using eight algorithms of feature selection (H, RE, ASA, OP, ACH, ACC, QSO, and the sequence order coupling number of each sequence) based on 9 amino acid windows size. They specifically showed which features are the most important ones and have more effects on accuracy. It was mentioned that AAC was the best feature for S and T sites. Then these features are used as RF input with 10-fold-cross validation. The accuracy of the model is approximately 80% for S, T, and Y sites [118]. Moreover, in the RF-phos-2.0, their RF model has improved by using window sizes of 5 to 21 amino acids and using different features. QSO was the best feature for S and T [98]. RF-phos-1.0 and RF-phos-2.0 specifically predicted universal p-sites. It should be mentioned that feature selection methods helped to improve the accuracy of various approaches.

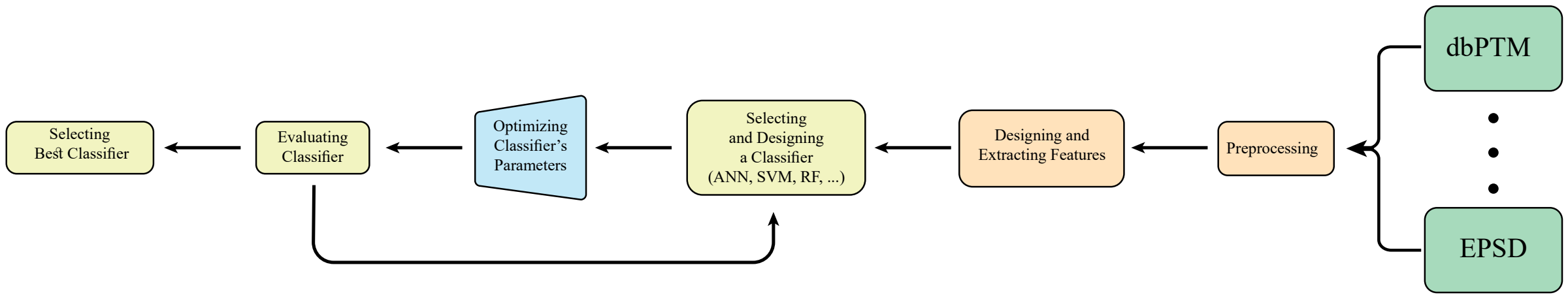

Fig 8: Procedure of using conventional machine learning methods.

Microbial Phosphorylation Site predictor (MPsites) was proposed to recognize universal microbial p-sites with different kinds of sequence features. In order to convert each sequence to numerical vectors, they used various sequence encoding strategies including AF, BE, AAINDEX, PWAA. They used naïve bayes, SVM, neural networks, decision tree and RF algorithms to recognize S and T p-sites. Results showed that RF has better performance than the other algorithms. It got 68% accuracy for S sites and 75% accuracy for T sites [129].

Cao etal. [130] proposed a method to predict phosphorylation in 7 species specific fungi proteins. They used two step feature optimizations to select important features and whichever is improved their SVM prediction performance model. KNN, Amino Acid Composition, di-Amino Acid Composition (AAC) and physiochemical properties (PCP) were used as features. First with RF model they ranked each input feature according to the mean accuracy. In the second step, top ten features from step one were merged with the training set to train the SVM model. In each step one feature was added to the model to improve the performance. The accuracy of their SVM model is over 80%.

Chen et al. [131] proposed a feature selection method named GAS, based on ant colony, genetic algorithm and evaluation strategies for six kinases types to choose the best classifier.

In a research Qui et al. [105] developed an approach called iPhos-PseEvo. Protein sequence evolutionary and pseudo amino acid composition (PseAAC) was selected as feature for ensemble RF model. The accuracy for their model was 71% with the jackknife test evaluation approach.

Furthermore, Multi-iPPseEvo [104] is similar to iPhos-PseEvo but with a different implementation method and using k-fold cross validation. This method contains a multi-ensemble RF classifier for each S, T and Y site and proposed multi-label p-site prediction for each S, T and Y site.

### 5.2.3 End-to-End learning approach

End-to-End learning becomes a hot topic in the ML field by taking the advantage of DL. DNN for short, is almost the same as the previous Artificial Neural Networks (ANNs) with minor modifications to be more effective and practical in representation learning. Similar to the human brain, each DNN's layer (or group of layers) could be used for learning the hierarchical abstraction for downstream tasks. In other words, usually raw input sequences (one-hot encoding) are just fed to a DNN and it does the feature selections inside of the layers by itself. Since it refers to training a possibly complex learning system by applying gradient-based learning to the system as a whole, it is called End-to-End learning. End-to-End learning systems are specially designed so that all modules are differentiable. In effect, all peripheral modules like representation learning and memory formation are covered by a holistic learning process [132]. Figure 9 shows the common procedure of End-to-End learning methods.

DL has made great success in solving problems that have resisted the best efforts of the AI community for years especially in different biological problems [133–139]. In recent years, there have been breakthroughs in DL which is the field applied to PTM classification especially protein phosphorylation prediction. Generally, these architectures are used for feature extraction and classification tasks at the same time. DL methods are multi-level representation learning methods that are achieved by stacking simple but non-linear layers. As mentioned earlier, the main aspect of this approach compared to the conventional ML approach is that the layer of features or the feature extraction step are not designed by human engineers or manually. These layers are acquired from input data to extract best patterns accurately and quickly. Though, the most important point about DL is that it needs huge amounts of data and in fact, by increasing the size of dataset it can perform better. This can be counted as a drawback by the way; when the dataset is not big enough, it quickly falls behind other ML methods in terms of performance.

Among all DL architectures, Convolutional Neural Networks (CNN), Recurrent Neural Networks (RNN) and Long Short Memory (LSTM) are the famous ones. Specifically, CNN has attracted more attention in PTM field [72,73,140]. MusiteDeep [73] provided a DL architecture called MusiteDeep to predict general and kinases-specific families' places. They used the window size of 33 amino acids for input sequences and presented multi-layer CNN and attention layers architecture. Sequences of proteins containing phosphorylated or non-phosphorylated sites are considered as binary classification problems. In DeepPhos' paper [72], in contrast to multi-layer models of MusiteDeep, they used dense CNN blocks that can show different representations of sequences. They called it DeepPhos which could improve the performance of MusiteDeep using different window sizes of 15, 33 and 51. Both these two methods DeepPhos and MusiteDeep are made for kinases family and universal p-sites. Moreover, the PhosTransfer [140] is a DL based framework which constructs pre-train architecture with CNNs based on kinases hierarchy and transfer learning. They believed that protein kinases within the same subfamily, family and group probably share similar local sequential and structural patterns and with this presumption they used pre-train feature extractions for fine tuning for lower levels of kinases. The Phos Transfer can achieve 0.89 AUC scores on average.

The DeepPPSite is a DL model based for predicting universal p-sites by considering the sequence information [74]. They used stacked LSTM architecture with one hot encoding, PSPM, EBGW, CKSAAP, and AAINDEX input features that could achieve 0.358 MCC value for S, 0.356 for T and 0.350 for Y p-sites.

Furthermore, there has been some works such as [141] which used hybrid architectures. they presented a specific End-to-End CNN-LSTM architecture and called it DeepIPs, to accurately predict universal p-sites in host cells infected with SARS-CoV-2 [142,143]. They utilize two approaches in Natural Language Processing as word embedding layers. First is Supervised Embedding Layers and the second one is Unsupervised Embedding Layers based on the Glove [144], the Fast Text [145,146] and the Word2vec [147] pre-train word embedding methods. 80.45 for S/T and 75.22 for Y accuracy was achieved.

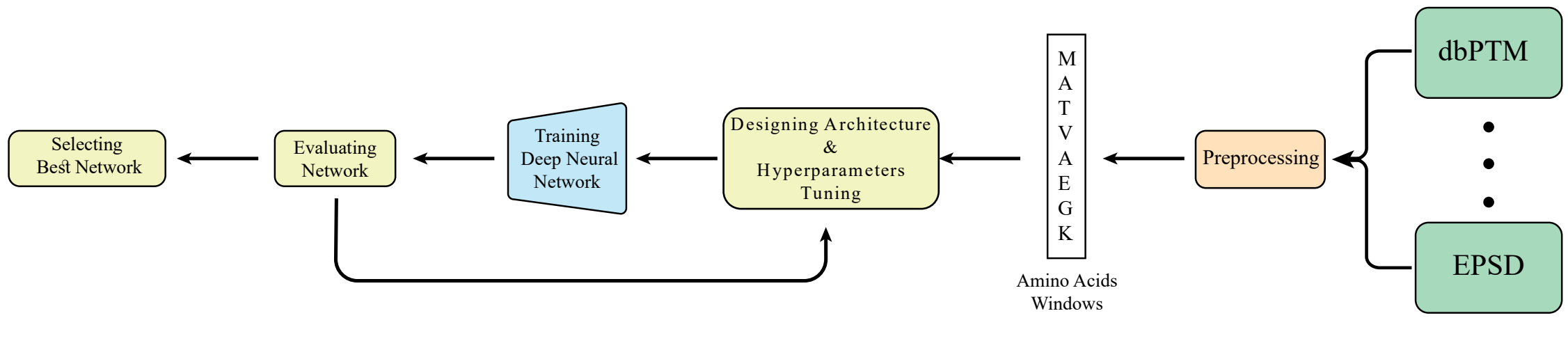

Fig 9: Procedure of End-to-End learning method.

DL provides a highly effective framework for dealing with modern-day learning challenges. The modern high-performance interpretable deep tabular learning network (TabNet) provides an extremely powerful framework for solving learning problems [148]. For example, Khalili et al. [77] developed a TabNet model to predict p-sites in soybean with a high accuracy rate that outperformed other common ML models (LR-L1, LR-L2, RF, SVM, and XGBoost). They assessed and compared the strength and reliability of all models using 10-fold cross-validation.

Experiments assessed the performance of AAC, DPC, TPC, PSSM, and physicochemical properties as individual features. To extract training sequences for model development, various window sizes ranging from 7 to 35 values were used. They got the best results for window size 13 with an accuracy of 87.34% based on PSSM features.

Naser and et al. [149] compared and analyzed human engineered representations and deep representations for the reorganization of Phosphoserine sites. They used DNN architecture like CNN, LSTM and RNN models for feature representation from sequences and compare their performance with human-engineered representations.

Even though, most DL approaches have built their architecture with large volumes of data, a paper [150] with small amount of data from only two kinases family and with considering windows with length of 9, proposed a simple DNN architecture. Their model has been able to achieve nearly 80% accuracy. It means that DL can also perform well on low data regions. This algorithm is designed for both kinases family and universal p-sites.

Guo and et. Al [151] collected phosphoprotein-binding domains (PPBD) that interact with phosphoprotein-binding domains containing proteins (PPCP) from 12 eukaryotic species and developed end to end DL method with transfer learning to classified Protein binding domains into a hierarchical structure with three levels, including group, family, and single PPBD cluster. They design 7 layers DNN network.

Despite most of End-to-End approaches uses raw sequences (one-hot encoding embedding), PhosIDN [152] used deep neural network with combining raw sequences and protein-protein interaction information together as inputs. Since this method was basically an End-to-End learning architecture with additional embedding input information, we chose it in this category. This architecture contains three sub-networks: a) sequence feature encoding sub-network (SFENet), b) PPI feature encoding sub-network (IFENet), c) heterogeneous feature combination sub-network (HFCNet).

Table 2: Introduces models and tool for p-sites proteins prediction. *: Refers to define general type or kinases type of proteins, U for universal sets and K for kinases family. **: Was not available at the time of writing.

| | **Acronym** | **Type / Description** | **Method** | **Feature extraction method** | **Dataset size** | **Window size** | **Redundancy threshold** | **Evaluation strategies** | **URL** | **U/K *** |
|---|---|---|---|---|---|---|---|---|---|---|
| Phosphorylation methods and online tools | NetPhos [125] | Conventional | ANN | Sequence composition features | 902 p-sites | 21 (Y, S) 25 (T) | - | 5-fold | http://www.cbs.dtu.dk/services/NetPhos/ | K |
| | [120] | Conventional | SVM | - | 855 S, 216 T | 3-25 | 70% | 7-fold | http://www.ngri.re.kr/proteo/PredPhospho.html ** | K |
| | [127] | Conventional | RF | Auto covariance transform, 7 physicochemical properties | 1,911 p-sites | - | 40% | 5-fold | ------ | K |
| | [109] | Conventional | SVM | EBAG, PWAA | 230 S, 61 T, 14 Y | 23 | - | 10-fold | ------ | U |
| | Rice-phospho 1.0 [111] | Conventional | RF | AF, CKSAAP, KNN | 4,220 S, 605 T, 141 Y | 25 | - | 10-fold | http://bioinformatics.fafu.edu.cn/rice_phospho1.0 | U |
| | GSVM [97] | Conventional | SVM | KNN, AF, DF, ACH | ~50,000 P-sites | 13 | 30% | | ------ | U |
| | RF-phos-1.0 [118] | Conventional | RF | H, RE, ASA, OP, ACH, AAC, QSO | ~28,000 p-sites | 5 to 21 | 30% | 10-fold | ------ | U |
| | RF-phos-2.0 [98] | Conventional | RF | H, RE, IG, ASA, OP, ACH, AAC, QSO | ~28,000 p-sites | 5 to 21 | 30% | 10-fold | http://bcb.ncat.edu/RFPhos/ ** | U |
| | PhosTransfer [140] | Conventional | CNN | H, RE, DF, OP, ACH | ~10,000 S, ~34,000 T, ~3,000 Y | - | 40% | - | https://github.com/yxu132/PhosTransfer | K |
| | deepPsites [74] | Conventional | LSTM | CKSAAP, EBGW, IPCP, PSPM | ~7,000 S, ~2,000 T, ~700 Y | 15, 19, 21 | 30% | 10-fold | https://github.com/saeed344/DeepPPSite | U |
| | GPS 5.0 [31] | Conventional | LR | Structural features | ~15,000 p-sites | 20 | - | 10-fold | http://gps.biocuckoo.cn | K |
| | MPSite [129] | Conventional | RF | AF, IP, PSSM, PWAA, SSF | ~2,700 S, 2,100 T | 7 to 25 | 30% | 10-fold | http://kurata14.bio.kyutech.ac.jp/MPSite/ | U |
| | Quokka [30] | Conventional | LR | KNN, AF, BLOUSM64 | ~2,400 S, ~370 T | 15, 19, 21 | 30% | 5-fold | http://quokka.erc.monash.edu/#webserver ** | K |
| | PhosContext2vec [153] | Conventional | SVM | H, BLOUSM64, DF, OP, ACH, Secondary structure | Universal: ~20,000 S, ~5,600 T, ~2,100 Y Kinases: ~4,100 | 25 | - | 10-fold | http://phoscontext2vec.erc.monash.edu/ | K/U |
| | PhosphoSVM [123] | Conventionlal | SVM | H, RE, Secondary structure, DF, ASA, OP, KNN, ACH | ~25,000 S, ~7,200 T, ~2,700 Y | 15, 19, 21 | 30% | 10-fold | http://sysbio.unl.edu/PhosphoSVM/ | U |
| | PhosPred-RF [154] | Conventional | RF | H, RE, IG, OP | ~4,300 S, ~2,700 T | 15, 19, 21 | 30% | 10-fold | http://bioinformatics.ustc.edu.cn/phos_pred/ ** | U |
| | [130] | Conventional | SVM | Sequence information, Evolutionary information, Physicochemical properties | Various for organisms | - | 30% | Independent test | http://computbiol.ncu.edu.cn/PreSSFP ** | U |
| | [131] | Conventional | Multiple classifiers | GAS | ~3,400 p-sites | - | - | 5-fold | ----- | K |
| | iPhos-PseEvo [105] | Conventional | Ensemble - RF | KNN, PseAAC | 845 S, 386 T, 249 Y | - | 50% | Jackknife test | http://www.jci-bioinfo.cn/iPhos-PseEvo ** | U |
| | Multi-iPPseEvo [104] | Conventional | RF | KNN, PseAAC | 845 S, 386 T, 249 Y | - | 50% | 5-fold | http://www.jci-bioinfo.cn/Multi-iPPseEvo ** | U |
| | deepIPs [141] | End-to-End | CNN-LSTM | - | 5,387 S/T, 102 Y | 33 | 30% | Independent test | https://github.com/linDinggroup/DeepIPs. ** http://lin-group.cn/server/DeepIPs/ | U |
| | DeepPhos [72] | End-to-End | CNN | - | 140,000 S/T, 27,000 Y | 15, 33, 51 | 40% | 10-fold | https://github.com/USTCHIlab/DeepPhos ** | U/K |
| | MusitDeep [73] | End-to-End | CNN + attention | - | ~35,000 S/T, ~2,000 Y | 33 | 50% | 5-fold | https://www.musite.net/ https://github.com/duolinwang/MusiteDeep_web | U/K |
| | [150] | End-to-End | DNN | - | ~1,800 S, 700 T, 200 Y | 9 | 20% | 10-fold | ------- | U/K |
| | PhosIDN [152] | End-to-End | SFENet+ IFENet+ HFCNet | PPI graph embedding | ~160,000 p-sites | 15, 33, 71 | 40% | Independent test | https://github.com/ustchangyuanyang/PhosIDN | U/K |
| | [77] | Neural network + feature | TabNet | AAC, DPC, TPC, PSSM, physicochemical properties | ~ 4 500 p-sites | 7 to 35 | 40% | 10-fold | ------- | U |

## 5.3 Tools for protein phosphorylation prediction

Due to the high cost and low speed of using experimental methods to recognize p-sites, in recent years many computational online tools have been developed to help and increase the quality of p-sites prediction. Table 2 introduced famous publicly accessible online tools or GitHub repositories for p-sites prediction.

# 6 Current limitations

In contrast to many ML domains, considering p-sites prediction methods on different datasets with different preprocessing, as well as different splitting in the train set and test set, it is not easy to accurately compare all of them and choose the best method. Therefore, we tried to evaluate some tools together by creating three new test datasets.

For this purpose, we selected the 2022 released version of dbPTM [61] database and picked up all new phospho-proteins in all organisms which did not exist in the previous versions. Subsequently, we built following test sets:

**161-all**: 161 new proteins with p-sites which were selected randomly from 161 new released organisms' proteins (One protein per organism). This test set consists of 13,403 sites which 402 of them were p-sites. The maximum and minimum length of sequences were 7,096 and 49 respectively.

**161-humans**: 161 proteins with p-sites which were selected randomly from new released homo sapiens' proteins. This test set consists of 7,383 sites which 714 of them were p-sites. The maximum and minimum length of sequences were 921 and 714 respectively.

**100-top**: 100 new proteins with p-sites from top 10 organisms which have the biggest new protein numbers (Ten proteins per organism). They were selected randomly. This test set consists of 9,321 sites which 507 of them were p-sites. The maximum and minimum length of sequences were 3,498 and 102 respectively.

Table 3 Evaluating of tools on three 161-all, 161-human and 100-top test sets.

| **Tool** | MusiteDeep [73] | | | PhosIDN [152] | | | NetPhos [125] | | |
|---|---|---|---|---|---|---|---|---|---|
| **Test set** | 161-all | 161-humans | 100-top | 161-all | 161-humans | 100-top | 161-all | 161-humans | 100-top |
| **TP** | 168 | 194 | 150 | 249 | 308 | 297 | - | 447 | 339 |
| **FP** | 1656 | 745 | 1044 | 4356 | 1140 | 2597 | - | 3701 | 5349 |
| **TN** | 11378 | 5927 | 7781 | 8678 | 5532 | 6228 | - | 2971 | 3476 |
| **FN** | 201 | 517 | 346 | 120 | 403 | 199 | - | 264 | 157 |
| **Accuracy (%)** | 86.14 | 82.91 | 85.09 | 66.60 | 79.10 | 70.00 | - | 46.30 | 40.93 |
| **Precision** | 0.09 | 0.21 | 0.13 | 0.05 | 0.21 | 0.1 | - | 0.11 | 0.06 |
| **Recall** | 0.46 | 0.27 | 0.3 | 0.67 | 0.43 | 0.6 | - | 0.63 | 0.68 |
| **F1** | 0.15 | 0.24 | 0.18 | 0.1 | 0.29 | 0.18 | - | 0.18 | 0.11 |
| **Specificity** | 0.87 | 0.89 | 0.88 | 0.67 | 0.83 | 0.71 | - | 0.45 | 0.39 |

Next, we tried to evaluate all universal p-sites prediction tools which were introduced on the above datasets. However, there were many hurdles in the evaluation stage. Kim et al. [120], RF-phos-2.0, PhosPred-RF, Cao et al. [130], iPhos-PseEvo, Multi-iPPseEvo were not available. Moreover, Rice-phospho 1.0 and PhosphoSVM only take one sequence as input in order to process and since the process was time consuming, we could not evaluate our three test datasets on them. Furthermore, DeepIPs did not have any response to our request. Finally, we selected three tools MusiteDeep, PhosIDN and NetPhos to evaluate. By the way, NetPhos could not predict sequences with length more than 4,000 amino acids and since the 161-all test set had proteins more than that length, we could not evaluate it. Table 3 shows the results.

As table 3 shows, all three tools performed weakly compared to what they reported on their papers. We interpreted from the results that there are not valid benchmarks for p-sites prediction. In other words, every paper and new method usually created a unique and different test set in order to report their method on it, which made it difficult to compare different methods together. Thus, for the fair and precise competition, we suggest that preparing uniform, comprehensive, unique and well-defined test benchmarks for p-sites prediction will be considered as a crucial step for the future research of this field.

## 7 Conclusion

Almost all proteins contain phosphorylation, which is responsible for critical functions in the cell. Various diseases can be caused by disruptions of this modification. So, this caused phosphorylation an important PTM type. The discovery of phosphorylation by high-throughput experimental methods is labor-intensive and time-consuming. Therefore, it's important to have powerful methods and tools to predict phosphorylation. As we investigated different review papers, we observed that there is not a complete review paper for p-sites predictions based on ML algorithms. For this reason, this paper briefly introduced some popular databases consisting of general PTM and specific type phosphorylation. Moreover, we introduced two important databases, EPSD and dbPTM, and compared them in order to analyze their distribution of p-sites.

Furthermore, we have given a brief overview of protein p-sites prediction by ML. In fact, ML approaches are mainly divided into classical methods and End-to-End learning methods. In addition to ML, we slightly discussed computational methods as well. Computational methods have statistical basis which are slow and have high time complexity. On the other hand, ML algorithms which are quite popular these days have attracted a lot of attention to use in p-sites prediction including SVM, LR, and RF. In conventional methods, SVM has shown better performance, although, it is clear that the feature extraction step would have a significant impact on the final result. Therefore, this study introduced 20 important and most used feature extraction methods based on the structural level, sequential, evolutionary based and physicochemical property-based categories. In a contrasting manner, CNN and RNN based architectures which have been famous in End-to-End learning, can predict p-sites directly from the raw input sequences without any feature extraction step. Consequently, researchers who turned to DL's End-to-End approaches, have reason to believe that feature extraction methods are time-consuming and need expert knowledge but the End-to-End ones do not require specialized knowledge.

In the next stage, evaluate the methods by different metrics for predicting p-sites approaches were reported to give standard metrics for comparison. Finally, in order to demonstrate current limitation in p-sites prediction methods, we created three test sets and evaluated available online tools on them. All those methods performed poorly compared to what they have reported in their papers which shows the importance of creating uniform and well-defined benchmarks for p-sites prediction.

# Acknowledgement

We gratefully acknowledge Hadi Pourmirzaei and Mohammad Ezati for preparing pictures and helps and recommendations.